\documentclass[twocolumn,pre,floatfix,superscriptaddress,showpacs]{revtex4}
\usepackage{dcolumn}
\usepackage{graphicx}
\usepackage{graphicx,amssymb,amsmath}
\usepackage{natbib}
\usepackage{color}
\usepackage{verbatim}
\usepackage[normalem]{ulem}
\usepackage{hyperref}

\begin{document}

%%%%%%%%%%%%%%%%%%%%%%%%%%%%%%%%%%%%%%%%%%%%%%%%%%%%%%%%%%%%%%%
%  This is version resubmission-final
%  Last Changed by AJL on Oct 15, 2013
%
%%%%%%%%%%%%%%%%%%%%%%%%%%%%%%%%%%%%%%%%%%%%%%%%%%%%%%%%%%%%%%%

\title{Shape transitions in soft spheres regulated by elasticity}

\author{Craig Fogle}
\affiliation{Department of Physics, UCLA, Los Angeles, CA 90095-1596,USA}

\author{Amy C.~Rowat}
\affiliation{Department of Integrative Biology and Physiology, UCLA, Los Angeles, CA 90095}
\affiliation{Department of Bioengineering, UCLA, Los Angeles, CA 90095, USA}

\author{Alex J.~Levine}
\affiliation{Department of Physics, UCLA, Los Angeles, CA 90095-1596,USA}
\affiliation{Department of Chemistry \& Biochemistry,UCLA, Los Angeles, CA 90095-1596, USA}
\affiliation{Department of Biomathematics, UCLA, Los Angeles, CA 90095-1596, USA}

\author{Joseph Rudnick}
\affiliation{Department of Physics, UCLA, Los Angeles, CA 90095-1596,USA}

\date{\today}

\begin{abstract}
Version 1 Resubmission. Last changed by AJL on Oct. 3\\
We study elasticity-driven morphological transitions of soft spherical core
shell structures in which the core can be treated as an isotropic elastic
continuum and the surface or shell as a tensionless liquid layer, whose elastic
response is dominated by bending. To generate the transitions, we consider the
case where the surface area of the liquid layer is increased for a fixed amount
of interior elastic material. We find that generically there is a critical
excess surface area at which the isotropic sphere becomes unstable to buckling.
At this point it adopts a lower symmetry wrinkled structure that can be
described by a spherical harmonic deformation.  We study the dependence of the
buckled sphere and critical excess area of the transition on the elastic
parameters and size of the system. We also relate our results to recent
experiments on the wrinkling of gel-filled vesicles as their interior volume is
reduced. The theory may have broader applications to a variety of related
structures from the macroscopic to the microscopic, including the wrinkling of
dried peas, raisins, as well as the cell nucleus.
\end{abstract}

\maketitle

\section{Introduction}

Euler buckling is a well-known mechanical instability. For a beam of length $L$
under axial compressive loads, there is a critical compressive strain $\sim
L^{-2}$ at which its deformation switches discontinuously from axial compression
to the lateral deflection~\cite{Euler1736,landau_elasticity}.  This mechanical
instability and the breaking of the axial symmetry of the compressed beam by its
lateral deflection provides a mechanical analogy to symmetry-breaking first
order phase transitions~\cite{Golubovic1998}. The role of Euler buckling in
beams has, of course, been explored extensively in engineering and materials
science~\cite{Hartog1977}, but it also plays an important role in biological
physics in the context of buckling cytoskeletal filaments where such buckling
affects force propagation along individual filaments~\cite{Das2008}, and the
nonlinear mechanics~\cite{Chaudhuri2007,Conti2009} and structural evolution
under load~\cite{Silva2011} of their networks.

Similar buckling and wrinkling instabilities appear in elastic
sheets~\cite{Nelson1987,Cerda2003} under in-plane compressive stress. In cases
where that compressive stress is generated by the growth or addition of material
to a constrained system, morphological transitions induced by buckling
result~\cite{Rodriguez1994,Drasdo2000,Goriely2005,Dervaux2008}, and these are
believed to have biological relevance in a variety of contexts including the
formation of fingerprints~\cite{Kucken2004} and intestinal structures -- villi
and crypts -- that can be reproduced {\em in vitro}~\cite{Sato2009} and
understood theoretically~\cite{Hannezo2011} by considering the buckling of a
thin elastic and growing layer coupled to a thick elastic substrate. More
generally, the appearance of wrinkling of a stiff thin elastic layer
mechanically coupled to a soft, three-dimensional elastic substrate has been
studied quite generally~\cite{Genzer2006,Yoon2010,Breid2011,Trejo2013} as has
that of thin elastic layers coupled to a fluid
substrate~\cite{Huang2006,Diamant2011}. The modifications of the buckling
phenomenon arising from coupling the elastic sheet to the substrate are
generically two-fold. First the coupling increases the critical load at which
buckling first occurs, and second it modifies the wavelength of the buckled
state near to the transition. The latter effect can be understood by noting that
longer wavelength buckling lowers the elastic energy of the thin sheet, but
generates deformations over longer length scales within the substrate at the
cost of greater elastic energy there. The competition between these two elastic
energies associated with the sheet and the substrate generate  minimum energy
buckled state at some intermediate wavelength. Without that competition, the
buckling wavelength is controlled entirely by the system size, as it is in the
archetypal case of Euler buckling in beam.

Morphological transitions can also be induced in compact elastic bodies,
including a variety of core-shell structured spheres. In this article, we
explore the surface wrinkling of a spherical elastic object bounded by a
membrane, which supports bending stresses.  Motivated primarily by the
observation surface buckling of a giant unilamellar vesicle filled with an
elastic gel as their interior gel was osmotically shrunk~\cite{Viallat2004}, we
examine theoretically morphological transition driving by an increase in the
surface area of the membrane-bounded sphere for fixed interior content. For
sufficiently small amounts of excess area, the system simply expands
isotropically. At a critical value of the excess area, however, the sphere
buckles into a lower symmetry shape. We have determine the critical excess area
and optimally  buckled shape within a linear elastic analysis. More broadly,
these results should be relevant to morphological transitions in a number of
elastic core-shell structures~\cite{Li2011} and even the cell
nucleus~\cite{Rowat2006}.

The remainder of the article is organized as follows. In section
\ref{sec:heuristic} we first present the results of a heuristic model replacing
the interior elastic deformation field with a scalar Laplacian model. This
theory provides insight into the single control parameter for the morphological
transition and the basic competition between the elastic energy cost of surface
deformation favoring buckling at longer wavelengths and interior strain favoring
buckling at shorter wavelengths.  In section \ref{sec:full-one} we present the
full elastic calculation. In section \ref{sec:results} we examine the solution
and analyze the first order buckling transition and determine the dominant
deformation mode in the buckled state. We find a first order transition to a
buckled state identified as a spherical harmonic mode $Y_{\ell m}$ with $\ell
\sim {\cal O}(10)$ for a broad range of the relevant elastic parameters and $ m
= \ell$, although there are typically generated $m$ states for larger $\ell$.
Finally, we conclude in section \ref{sec:summary} with a discussion of the
implications of these results to current experiments and proposals for future
tests. A few details of a more technical nature are relegated to the appendix.

\section{The model}

\subsection{Scalar Model}
\label{sec:heuristic}

It is helpful to first examine a heuristic model that dramatically simplifies
the elastic calculation of the interior of the sphere, but still manages to
capture the basic competition between surface and bulk elastic energies.  To
that end we replace the vectorial elastic displacement field ${\bf u}({\bf x})$
by a scalar field $\phi({\bf x})$ that obeys the Laplace equation rather than
the usual equation of stress continuity in a bulk elastic medium. In this way we
neglect the distinction between transverse and longitudinal modes of the solid,
but retain the basic power counting of derivatives in the equation of stress
continuity, which is necessary for our argument.  We impose ``excess area'' on
the surface of the elastic sphere of radius $R$ by requiring the scalarized
deformation to be given there by
\begin{equation}
\label{scalar-boundary}
\phi(r=R,\theta) = \Phi P_{\ell}(\cos \theta),
\end{equation}
where $P_{\ell}$ is a Legendre polynomial and $\theta$ is the polar angle. In
this way we impose a deformation on the surface in the form of a spherical
harmonic. In the scalarized approach we specialize to azimuthally symmetric
modes ($m=0$) but this is not essential and is relaxed in the full calculation
presented below. The surface deformation can be thought of as being imposed by a
set of normal stresses on the sphere's surface that serve as a method for
injecting excess area on the surface. The morphological transition can be
understood as resulting from either imposing excess surface area or reducing the
interior volume of the unstrained material. For example, shape transitions are
observed when  gel-filled vesicles are osmotically shrunk~\cite{Viallat2004}. In
either case, the application of surface stresses is equivalent to changing the
reference state within linear elasticity theory -- see appendix A for a more
detailed discussion.

Given that $\phi({\bf x})$ satisfies the Laplace equation, a solution satisfying
the above condition is given in terms of the polar angle $\theta$ and radial
distance $r$ from the origin of the coordinate system at the center of the
sphere by

\begin{equation}
\label{scalar-solution}
\phi(r,\theta) = \Phi \left( \frac{r}{R} \right)^{\ell}  P_{\ell}(\cos \theta).
\end{equation}
The deformation energy of the elastic interior is
\begin{equation}
\label{scalar-interior-energy}
E_{\rm bulk}= \frac{\gamma}{2} \int_{r < R} d^{3}{\bf x} \left( \nabla \phi \right)^{2},
\end{equation}
where $\gamma$ is the single elastic constant in the scalarized elastic model,
which has dimensions of energy per length. Inserting the solution
Eq.~\ref{scalar-solution} into Eq.~\ref{scalar-interior-energy}, we find that
\begin{equation}
\label{scalar-bulk-energy}
E_{\rm bulk} = 2 \pi \gamma R \Phi^{2} \left[ \frac{1 + \ell (\ell+1)}{(2 \ell+1)^{2}}\right].
\end{equation}
The prefactors associated with the elastic constant, sphere's radius, and
magnitude of deformation follow naturally from dimensional analysis; the
%%%%%%%%%%  FIGURE
\begin{figure}[h]
 \centering
\includegraphics[width=3.375in]{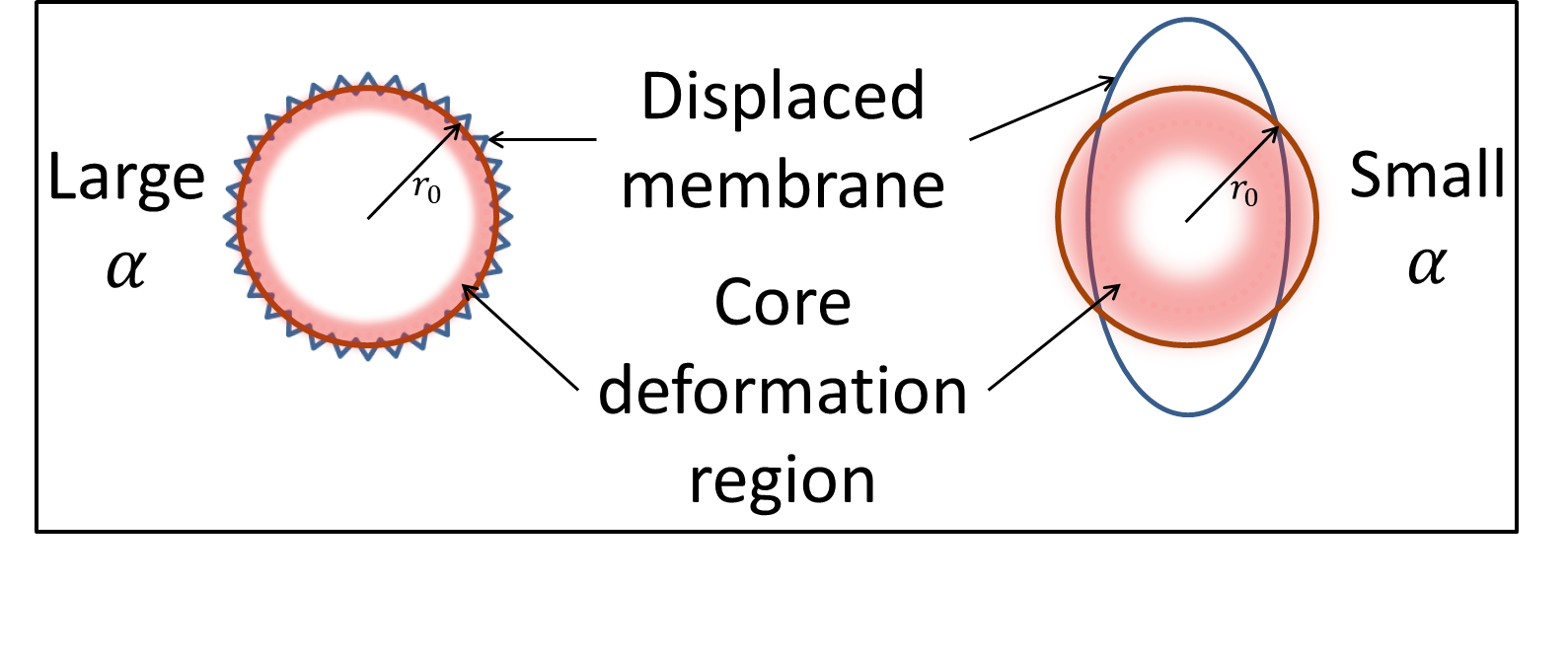}
\caption{
A schematic representation of the basic energy balance between the surface and
the bulk. By inserting excess surface area in a short wavelength
deformation as shown on the left, the deformation field in the interior decays
rapidly, minimizing bulk elastic energy at the cost of the incurring high
bending energy on the surface. Inserting the same excess area into a lower
spherical harmonic, as shown on the right, exchanges the relative importance of
surface and bulk energetic contributions. This trade off between bulk and
surface energy selects a optimally buckled state with
$\ell \sim \alpha^{1/3}$. See text for details.}
\label{fig:schematic}
\end{figure}
%%%%%%%%%%%%%
$\ell$-dependent term in square brackets, however, shows that higher order
spherical harmonics (of the same amplitude) generate less elastic energy storage
in the interior. In terms of the energy of the $\ell =0$ mode $E_{0}= 2 \pi
\gamma R \Phi^{2}$, the elastic energy associated with the $\ell^{\rm th}$ mode
decays monotonically with $\ell$ to the limiting value of $E_{\rm bulk}
\rightarrow  \tfrac{E_{0}}{4}$ as $\ell \rightarrow \infty$.

We now introduce a bending energy associated with the surface deformation
imposed by $\phi({\bf x})$, specified in Eq.~\ref{scalar-boundary}.  Letting
$\nabla_{\perp}$ be the parts of the gradient operator normal to the radial
direction, we approximate this bending energy as
\begin{equation}
\label{scalar-bending-energy}
E_{\rm surf} = \frac{\tilde{\kappa}}{2} R^{2} \oint_{{r=R}} d \Omega
\left[ \nabla_{\perp}^{2} \phi \right]^{2},
\end{equation}
where $\tilde{\kappa}$ is the bending energy, which, in the scalarized model, has
dimensions of length squared times energy.  This form of the bending energy is
correct only for small deformations from an originally flat surface, so it is
approximate in this case and can only be justified {\em a posteriori} by showing
that energy minimizing buckling wavelength is small compared to the sphere's
circumference -- see below. In the full calculation, shown in the next section,
such approximations are not used. Using the solution Eq.~\ref{scalar-solution}
we find the bending energy to be given by
\begin{equation}
\label{scalar-surface-energy}
E_{\rm surf} = \frac{2 \tilde{\kappa} \pi \Phi^{2}}{R^{2}}
\left[ \frac{\ell^{2}(1+\ell)^{2}}{2 \ell + 1}\right].
\end{equation}
The sum of the energetic contributions from the bulk Eq.~\ref{scalar-bulk-energy}
and the surface Eq.~\ref{scalar-surface-energy} give the total energy
cost of a given deformation in mode $\ell$, $E_{\ell}$ . To determine the
minimum such energy for a given {\em excess surface area} $\Delta A$ we first
compute the ``excess area'' associated with the deformation $\phi$,
\begin{equation}
\label{scalar-excess-area}
\Delta A = \frac{R^{2}}{2} \oint d \Omega \left[ \nabla_{\perp} \phi \right]^{2},
\end{equation}
and use this result to eliminate the amplitude of the deformation field $\Phi$
in favor $\Delta A$.  Defining a dimensionless
energy ${\cal E}_{\ell} = R^{2}E_{\ell}/2 \pi \tilde{\kappa}$ we find that the
energy of a deformation in the $Y_{\ell}^0$ mode generating excess surface
area $\Delta A$ is given by
\begin{equation}
\label{scalar-total-energy}
{\cal E}_{\ell}= \frac{\Delta A}{\ell (\ell+1)}
\left[ \tilde{\alpha} \frac{1 + \ell (\ell+1) }{2 \ell +1} + \ell^{2} (\ell+1)^{2} \right],
\end{equation}
where we have introduced a single dimensionless control parameter
\begin{equation}
\label{scalar-alpha}
\tilde{\alpha} = \frac{\gamma R^{3}}{\tilde{\kappa}}.
\end{equation}
This dimensionless constant describes the relative stiffness of the interior
as compared to the surface.  For large $\ell$ one notes that the dimensionless
energy in Eq.~\ref{scalar-total-energy} is given roughly by
\begin{equation}
\label{scalar-energy-scaling}
\frac{{\cal E}_{\ell}}{\Delta A} \sim \frac{\tilde{\alpha}}{\ell} + \ell^{2},
\end{equation}
showing that a minimum exists for a mode at
\begin{equation}
\label{ell-star-scaling}
\tilde{\ell}^{\star} \sim \tilde{\alpha}^{1/3}.
\end{equation}

The key features determining the form of Eq.~\ref{scalar-energy-scaling} can be
readily understood. The $\tfrac{\tilde{ \alpha}}{\ell}$ and $\ell^2$ represent the
energetic contribution from the interior and the surface respectively. The
growth of ${\cal E}_{\ell}$ at large $\ell$ reflects the high bending energy
cost of the high $\ell$ modes, but the first term, weighted by the ratio of the
bulk to surface modulus, decreases with increasing $\ell$. The deformation field
in the sphere's interior becomes confined to a type of boundary layer: $\phi
\sim (r/R)^{\ell}$. Fig.~1 shows a schematic representation of this trade off
between the confinement of the interior strain to a boundary layer at high
$\ell$, leading to the reduction of the bulk energy and the monotonically
growing bending energy cost of
injecting a fixed amount of excess surface area into a mode with increasing $\ell$.

The qualitative features of this result will survive the introduction of a
proper vectorial displacement field to describe the deformation state of the
sphere. In that case we will replace the dimensionless control parameter $\tilde{\alpha}$ by
\begin{equation}
\label{alpha-definition}
\alpha = \frac{\mu R^{3}}{\kappa},
\end{equation}
where $\mu$ is the shear modulus of the bulk and $\kappa$ is the bending modulus
of the surface. Ignoring a weak dependence on the Poisson ratio of the bulk
elastic material, we will find a similar structure of the total energy yielding,
a transition from a spherical state to deformation in the mode $\ell^{\star}$.
The generic nature of the result is not surprising in light of previous work on
buckling of beams embedded in an elastic medium~\cite{Brangwynne2006}. In that case, the beam
is generically stiffened against buckling by the surrounding medium and the
unstable mode associated with the buckled state shifts from the lowest
wavelength allowed by the length of the beam to a  finite wavelength set by the
ratio of the beam's bending modulus and the elastic constants describing the
surrounding medium~\cite{Brangwynne2006}. The finite wavelength results from the
fact that the continuum elastic theory is Laplacian. Undulations of the beam at
wavevector $k$ generate strain in the elastic continuum that decays over a distance $1/k$
into the medium. This sets up the same trade off between beam bending energy and
bulk elastic energy as discussed here. The key differences in the present
calculation involve the spherical geometry of the undeformed structure, which
renders the calculation more difficult.

\subsection{Full Elastic Calculation}
\label{sec:full-one}

To calculate the energy associated with the membrane bounding the elastic
sphere we introduce the Helfrich energy of that surface that depends on
the mean curvature $H$, the spontaneous curvature $H_{s} =2/R_{s}$, and a
single elastic constant, the bending modulus $\kappa$:
\begin{equation}
\label{surface-energy}
E_{\rm surface} = \frac{\kappa}{2} \oint_{{r=R}} R^{2}d\Omega \left(H-H_s\right)^2,
\end{equation}
where the integral is over all solid angles. Since we consider only
genus-preserving morphological transitions, we may neglect terms proportional to
the Gaussian curvature modulus~\cite{Safran2003}.

To determine the elastic energy stored in the sphere's interior, we
parameterize the strain in the usual way by introducing the symmetrized
deformation tensor $u_{ij} = (1/2)[\partial_{i} u_{j} + \partial_{j}u_{i}]$
calculated from the displacement field ${\bf u}({\bf x})$ defined everywhere
within the undeformed body, a sphere of radius $R$. The elastic properties of the interior
 are described in terms its shear modulus $\mu>0$ and the Poisson ratio $-1<\nu<1/2$,
with $\nu = 1/2$ corresponding to an incompressible solid. In terms of these
parameters the elastic energy associated with a given deformation
state of the sphere is given by~\cite{landau_elasticity}
\begin{equation}
\label{bulk-energy}
E_{\rm bulk} = \mu \int_{|{\bf x}|<R} d^{3}x
		 \left(\frac{\nu}{1-2\nu} u_{ii}^2 + u_{ij}^{2} \right).
\end{equation}
The integral is over the reference space of the undeformed sphere. The above
strain tensor satisfies the standard stress continuity equation
$\partial_{i} \sigma_{ij}=0$ within the sphere requiring that ${\bf u}$ satisfy
\begin{equation}
\label{stress-continuity}
\frac{1}{1-2\nu} \nabla \nabla \cdot {\bf u}+ \nabla^2 {\bf u}=0
\end{equation}

The total energy of deformation, which is the sum of the contributions from
Eqs.~\ref{surface-energy},  \ref{bulk-energy}
is fully determined once one requires that the state of deformation of the
elastic interior  matches the imposed normal displacements at its boundary
surface,  $r=R$ . To do that we require the normal displacements obey
\begin{equation}
\label{boundary-condition-u}
{\bf u}(R, \theta, \phi) \cdot {\bf \hat{r}} = R g(\Omega),
\end{equation}
with $\Omega = (\theta,\phi)$ representing the polar and azimuthal angles.
Since we consider a fluid membrane boundary
and look at static states of deformation, the tangential displacements of the
elastic core ${\bf u}_{\perp}(r=R,\Omega)$, ${\bf u}_{\perp}\cdot \hat{r}=0$
do not affect the deformation energy of the surface. We take these displacements
to be zero for now and return to this point later in the discussion.

Following the analysis of the heuristic model above, we impose a
deformation on the surface of the sphere so that the radial distance $r(\Omega)$
from its center to the deformed boundary is expanded in spherical
harmonics~\cite{Milner1987}.  In other words
\begin{equation}
\label{deformed-sphere}
r(\Omega)=R \left[1+g(\Omega)\right]= R \left[1+\sum_{\ell,m} g_{\ell m} Y_{\ell m}(\Omega) \right].
\end{equation}
In order to make the radial deformation of the sphere explicitly real, we have introduced linear combinations of the complex spherical harmonics
$Y_{\ell}^{m}$ for $m>0$
\begin{equation}
\label{Ylm}
Y_{\ell m}= \tfrac{1}{\sqrt{2}}\left[Y_\ell^m+(-1)^mY_\ell^{-m}\right],
\end{equation}
and set $Y_{\ell 0} = Y_{\ell}^{0}$ for the azimuthally symmetric mode. Other linear combinations of the
spherical harmonics of the form $Y_{\ell}^{m}e^{i m \phi_{0}} + (-1)^{m}Y_{\ell}^{-m}e^{-i m \phi_{0}}$
correspond to the same deformation state, but simply rotated azimuthally by $\phi_{0}$. As such, they are not interesting,
at least until one considers the energy cost of linear combinations of such
deformation modes. Finally, it will prove convenient to rescale the amplitude the $\ell=0$ mode by introducing the notation $\bar{g}_{0} = g_{0}/\sqrt{4 \pi}$.

Since the surface energy Eq.~\ref{surface-energy} is quadratic in the deformations, the spherical harmonic decomposition of the surface deformation
energy yields a result quadratic in amplitudes $g_{\ell m}$ for $\ell \neq 0$.
\begin{eqnarray}
\nonumber
E_{\rm surface} &=& \bar{E}  +2\pi\kappa \bar{g}_0\omega_s \left[\omega_s(2+\bar{g}_0)-2\right] + \\
\nonumber
 & & +\frac{\kappa}{8} \sum_{\ell\neq 0,m} g_{\ell m}^2
 \left[\ell (\ell+2)(\ell^{2}-1) -4\omega_s\ell(\ell+1) \right.\\
& & \left. + 2 \omega_s^2(\ell^{2} + \ell +2)\right],
\label{surface-energy-decomposition}
\end{eqnarray}
where $\omega_{s} = R/R_{s}$ is the ratio of the undeformed sphere's radius to
the spontaneous curvature of its bounding membrane, and $\bar{E} = 2\pi\kappa
\left(1-\omega_s\right)^2$ is the energy associated with the mismatch
($\omega_{s}\neq 1$) between these quantities.

Such a mismatch $\omega_s\neq 1$ leads to a``prestress'' in the membrane-sphere system due
to the incompatibility of their undeformed states. We do not explore these effects here,
but it is clear that the prestress above can lead to buckling at zero excess area.

We may also calculate change in surface area associated with the deformation of
the sphere. Keeping terms to ${\cal O}(g^{2})$ we find
\begin{eqnarray}
\frac{A}{4 \pi R^{2}} &=&   \left(1 + \bar{g}_0 \right)^2+   \frac{1}{8 \pi R^{2}} \sum_{\ell\neq 0,m} g_{\ell m}^2
				\left[\ell^{2} + \ell + 2\right].
\label{surface-area-decomposition}
\end{eqnarray}
The first term is the extra area associated with the uniform change in the
sphere's radius, i.e., the $\ell =0$ mode. The second term gives the
contributions from all the higher modes of deformation $\ell >0$. Both the
results for the bending energy and contribution to the surface area are
consistent with previous work on the undulations of micelles~\cite{Milner1987}.
In the following we will consider the fractional excess area $\Delta$ of the
deformed sphere as the control parameter for the transition from spherical to
deformed shapes. Consequently, we define
\begin{equation}
\label{Delta-def}
\Delta = \frac{A}{4 \pi R^{2}} - 1.
\end{equation}
The surface energy Eq.~\ref{surface-energy-decomposition}, and the excess area
Eq.~\ref{surface-area-decomposition} both take the form of sums of terms each
proportional to the square of the spherical harmonic mode amplitudes. This will
also be true of the elastic energy stored in the sphere's interior Eq.~\ref
{bulk-energy}. Thus, in a deformation state given by a linear combination of
such modes, the contributions of each mode to both the energy cost and excess
surface area are simply additive. For this reason it is sufficient to consider a
single spherical harmonic mode at a time.

The remaining calculation of the energy stored in the elastic interior is
somewhat tedious; we present the key features of the calculation here and the
remaining details in appendix B.
Our method of solution has been used to consider the deformations of
spherical inclusions embedded in another elastic medium~\cite{Leo1985}, but we consider the
problem with displacement, rather than stress, boundary conditions at the surface.
To briefly describe the approach:
if one were to obtain a harmonic
vector ${\bf S}_{\ell m}({\bf x})$ i.e., $\nabla^{2} {\bf S} = 0$ that matches
the boundary conditions imposed on the displacement field at the boundary $r=R$
so that ${\bf S}(r=R,\Omega) = {\bf \hat{r}} g_{\ell m} Y_{\ell m}(\Omega)$,
then a displacement field of the form
\begin{equation}
\label{displacement-solution}
{\bf u}({\bf x}) = M_{\ell} (R^{2} - r^{2}) \nabla \nabla \cdot {\bf S}_{\ell m} + {\bf S}_{\ell m},
\end{equation}
with
\begin{equation}
M_{\ell} = \frac{1}{4 (1-\nu) (2 \ell -1) - 2 \ell}.
\end{equation}
is a solution to Eq.~\ref{stress-continuity} with the correct boundary
conditions, as may be checked by direct computation.

Recognizing that $Y_{\ell m}$ is a linear combination of complex spherical
harmonics $Y_{\ell}^{\pm m}$, it is sufficient to produce a complex harmonic
vector field ${\bf {\cal S}}$ that generates the appropriate part of the radial
displacement at the surface:
\begin{equation}
{\bf {\cal S}}(r=R,\Omega) = {\bf \hat{r}} Y_{\ell}^{m}(\Omega).
\end{equation}
The Cartesian components of such a complex vector are products of $\hat{r} \cdot
\hat{e}_{i}$, $ i = x,y,z$ and $Y_{\ell}^{m}$. Using the usual rules for adding
angular momentum, such products can be expanded in spherical harmonics having
$\ell' = \ell -1, \ell, \ell+1$ and $m' = m-1,m,m+1$ with coefficients given by
the Clebsh-Gordon coefficients.  The net result is a displacement field that
gives the appropriate radial displacement on the surface in a single $Y_{\ell m}$
mode, but contains a sum of terms including that mode as well as modes with $\ell \pm 1$ (and $m
\pm 1$) in the interior. Taking the necessary derivatives, one obtains the strain
tensor and, from that, the elastic energy density. Integrating the
energy density over the volume of the sphere, we obtain the total bulk elastic
energy. That result, combined with
Eqs.~\ref{surface-energy-decomposition}, \ref{surface-area-decomposition},
completes our analysis of the geometry and energetics of the sphere held in a
deformed state. As in the heuristic example, it is necessary only to eliminate
the amplitude of the surface deformation in favor of the true control variable,
the excess area (using Eq.~\ref{surface-area-decomposition}).
For a deformation consisting of a single harmonic with $\ell$ this
leads to
\begin{equation}
	g_{\ell m}(\Delta)=\left\{
	\begin{array}{ll}
		\sqrt{4\pi}\left(\sqrt{1+\Delta}-1\right) &\mbox{for }\ell=0 \\
		\sqrt{\Delta}\left(\frac{8\pi}{\ell(\ell+1)+2}\right)^{1/2}&
		\mbox{for }\ell \neq 0
	\end{array}\right. ,
	\label{amplitude-to-excess-area}
\end{equation}
where the fraction excess area $\Delta$ is defined by Eq.~\ref{Delta-def}.

For small $\Delta$, the amplitude of the $\ell =0$ mode grows as $g_{00} \sim \Delta$.
For all other modes with $\ell>0$ the amplitude grows as $g_{\ell m} \sim \Delta^{1/2}$.
Since the deformation energies of both the surface and the bulk are proportional to $g^2$
(when $\omega_s=1$, i.e., no prestress), the energy cost for inserting the the excess area
into the $\ell=0$ mode grows as $\Delta^2$. For the $\ell>0$ modes, however, it
grows linearly.
%%%%%%%%%%%  FIGURE
\begin{figure}[h]
\centering
\includegraphics[width=3.375in]{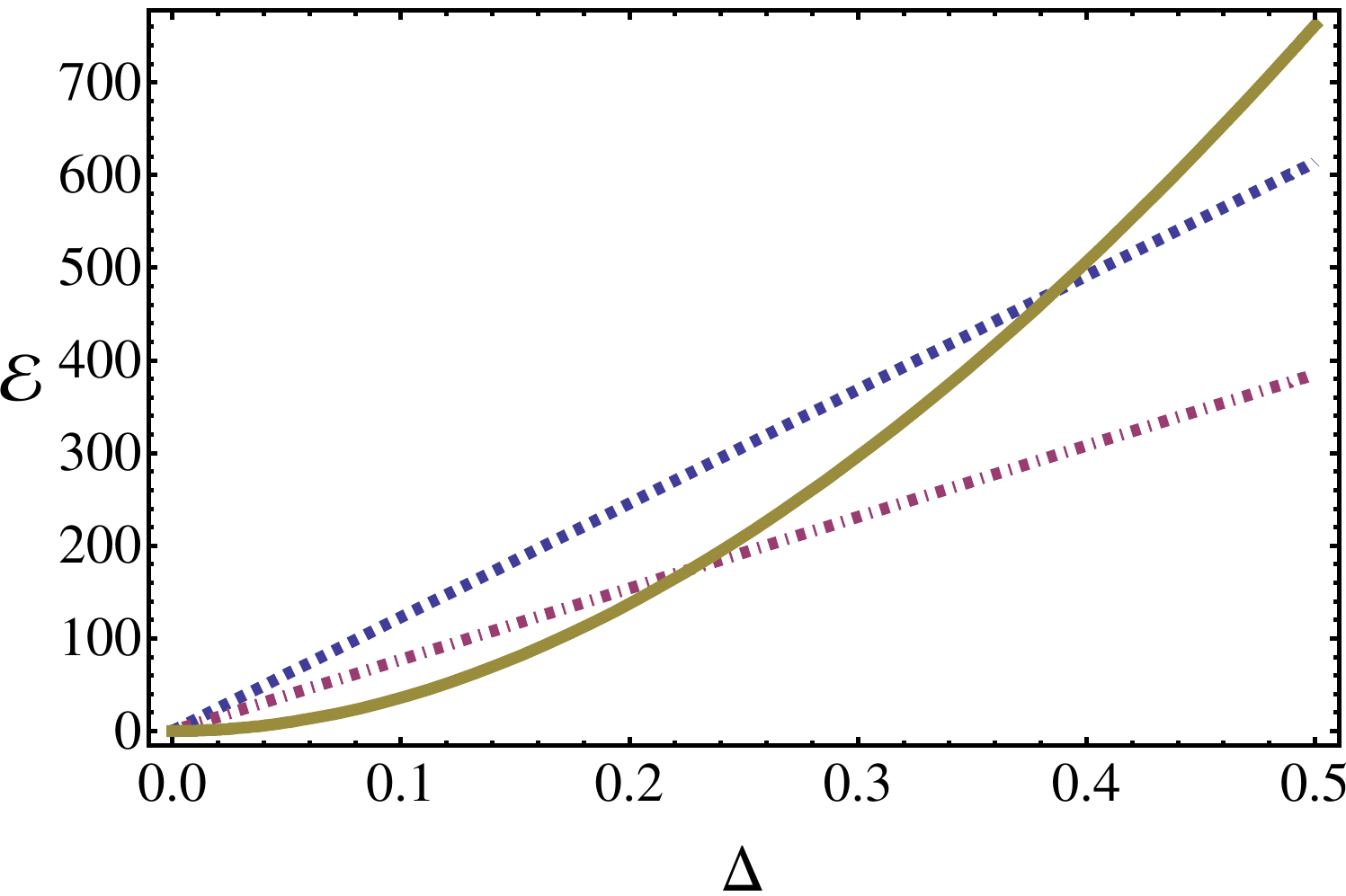}
\caption{(color online) The dimensionless energy ${\cal E} = E/\kappa$ of the deformed sphere and membrane as a function of the
fractional excess area $\Delta$ for
isotropic growth $\ell =0$ solid (gold) and for buckling into the $\ell =8, m = 0$ dashed (blue) mode or the  $\ell =8, m=8$ dot-dashed (purple).
The dimensionless control parameter is $\alpha = 575$
and the Poisson ratio of the interior is $\nu = 0.2$. As discussed in the text, the $\ell =8, m=8$ mode is the minimum energy buckled state of the sphere.
The transition from isotropic expansion to buckling
occurs at $\Delta \approx 0.22$.}
\label{fig:E_vs_D}
\end{figure}
%%%%%%%%%%%%%%%
Thus, for sufficiently small $\Delta$, the isotropic
deformation is necessarily lower in energy. At some $\ell$- and $m$-dependent
values of the excess area, the energetic cost storing that excess area in these
finite $\ell$ modes must become lower than storing it in the isotropic one. At
that point a morphological transition results. This is illustrated in
Fig.~\ref{fig:E_vs_D} which shows the nondimensionalized energy of the system
${\cal E} = E/\kappa$ as a function of the fractional excess area $\Delta$ --
see Eq.~\ref{Delta-def} -- in three cases: Accommodating that excess area by
isotropically expanding the sphere and membrane system (sold line, gold) or by
creating a lower symmetry buckled state in the $\ell =8, m=0$ (dashed line,
blue) or $\ell =8, m=8$ (dashed dotted line, purple) mode. It is clear that
there is a transition near $\Delta^\star \simeq 0.22$ where the isotropic solution is
no longer the energy minimizing solution.

As with the heuristic toy model, $\alpha$ -- see Eq.~\ref{alpha-definition} --
is the key dimensionless parameter controlling both the fractional critical
excess area of the transition $\Delta^{\star}$ and the symmetry of the optimally
buckled state above the transition $\ell^{\star}$. The Poisson ratio of the
elastic material inside the sphere has a weak effect on the quantitative results
as long as $\nu < 1/2$, i.e., as long as the material is not restrictly
incompressible. We return to this point in the next section. Of the three cases
shown, the optimal solution is the $\ell =8, m=8$ one; it will be shown later
that, for these parameters giving $\alpha = 575$ and $\nu =0.2$, there are in
fact six degenerate optimal solutions with $\ell =8$ and $m = 3,\dots,8$. In the
next section we survey the dependence of the buckling transition on $\alpha$,
examine the buckled state more closely, and present an overall phase diagram of
core/shell buckling.

\section{Results}
\label{sec:results}
\begin{figure}[h]
\centering
\includegraphics[width=3.375in]{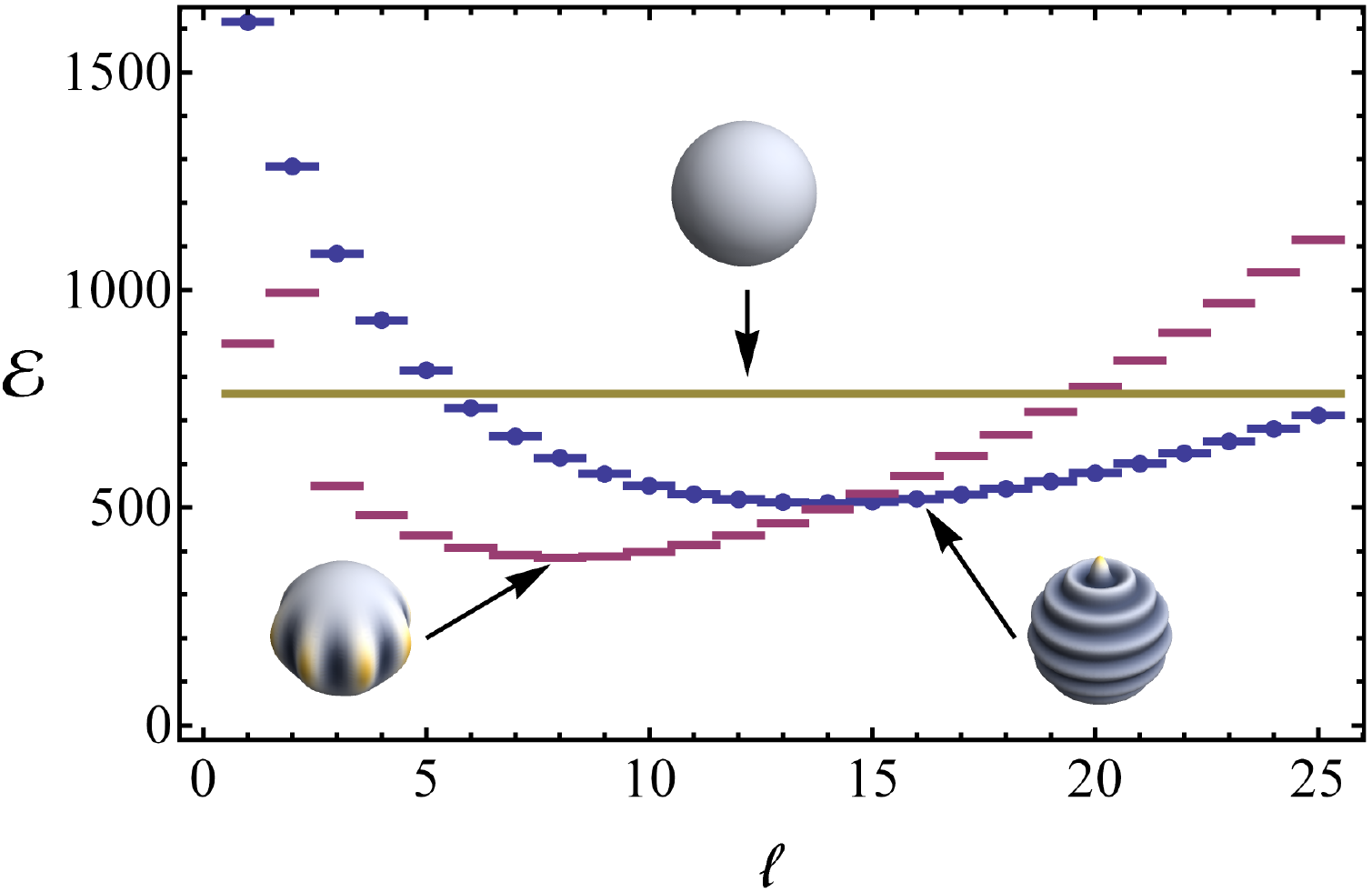}
\caption{(color online) Plot of the dimensionless energy ${\cal E} = E/\kappa$ vs.\ the angular harmonic of the buckled state $\ell$ for $\alpha = 575$, $\nu = 0.2$, and fixed
fractional excess area $\Delta = 0.5$. These plots can be used to determine the optimal buckled state symmetry $\ell^{\star}$.
The solid (gold) line shows the energy of the isotropically expanded state. The dashed (red) curve gives the energies of the $\ell =m$ state and the dot dashed
(blue) curve the energies of the $m=0$ state at each $\ell$.  For this $\alpha$, $\nu$ combination the lowest energy configuration at $\Delta =0.5$ is given by $\ell = m = 8$. Representative figures of
the isotropic state (top), the $\ell = m=8$ state (bottom left) and the $\ell =16, m=0$ (bottom right) are shown.}
\label{fig:E-vs-ell}
\end{figure}

\begin{figure}[h]
\centering
\includegraphics[width=3.375in]{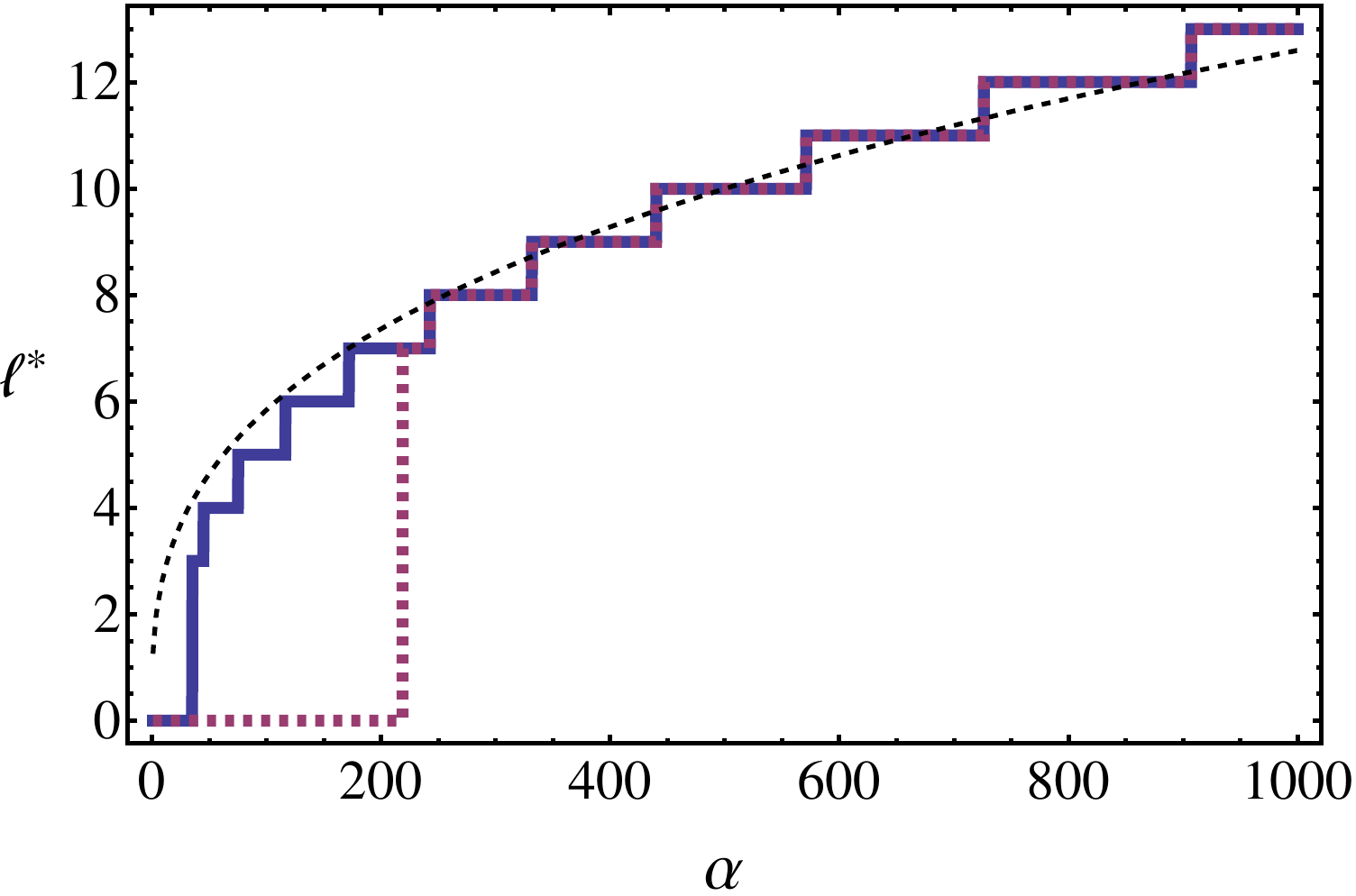}
\caption{(color online) Plot of $\ell^{\star}$, the angular harmonic of the minimum energy state of the sphere as a
function of $\alpha$ for fixed fractional excess areas $\Delta =0.25$ dashed (purple) and
$\Delta = 0.4$ solid (blue).  For sufficiently soft interiors (small $\alpha$) the isotropic shape is the energy minimum. As $\alpha$ is increased,
the sphere transitions to a buckled state with $\ell^{\star}$, and $m = \ell^{\star}$.  As expected from the heuristic treatment of scalarized elasticity, symmetry of the optimally buckled state
$\ell^{\star} \sim \alpha^{1/3}$, as shown by the dashed (black) line.}
\label{fig:lstar-vs-alpha}
\end{figure}

Having seen that a buckling transition is required from a basic scaling analysis, we examine in more detail which buckled configuration is an
energy minimum and which material parameters control the symmetry of the optimally buckled shape. To determine which is the optimally buckled shape
above the transition, i.e., for $\Delta > \Delta^{\star}$, we plot the dimensionless energy ${\cal E}$ as a function of the angular
harmonic $\ell$, using the same values of the material parameters as above ($\alpha =575$ and $\nu=0.2$).
A representative example is shown in Fig.~\ref{fig:E-vs-ell} in which the energy of the buckled sphere is plotted for buckling mode
$\ell, m=0$ (dashed dot, blue),  and $\ell, m = \ell$ (dashed, purple). For comparison, we show the energy of the isotropically expanded state as a solid (gold)
line. In all cases the fractional excess area is fixed at $\Delta = 0.5 > \Delta^{\star} \simeq 0.22$, so some lower symmetry state of the system is the
energy minimum. Here, we see that the $\ell^{\star} =8$ state with large $m$ is the lowest energy buckled state of the system.  We will find that the
larger $m$ solutions for a given $\ell$ are, in fact, generically of lower energy.
%%%%%%%%%%%%  FIGURE
%%%%%%%%%%%%%%%%
Repeating this exercise for various values of $\alpha$, one determines both the
dependence of the critical fractional excess area $\Delta^{\star}$ and the
symmetry of the minimum energy buckled state $\ell^{\star}$ on the control
parameter $\alpha$.  The results for the dependence of $\ell^{\star}$ upon
$\alpha$ are shown in Fig.~\ref{fig:lstar-vs-alpha} for two different values of
the fractional excess area: $\Delta = 0.4$ dashed (blue)  and $\Delta = 0.25$
solid (red).  As expected based on the heuristic model, we find that
$\ell^{\star} \sim \alpha^{1/3}$, as shown by a comparison to the cube root
function dashed by the dashed line in the figure. For sufficiently small values
of $\alpha$ the interior of the sphere is too compliant to drive the buckling
transition; the isotropic deformation of the sphere is the energetically favored
manner of storing the excess surface area. Collecting more results of this form,
we observe that the critical fractional excess area decreases with increasing
$\alpha$ for fixed Poisson ratio as $\Delta^{\star} \sim \alpha^{-1/3}$, as is
discussed later in terms of the overall buckling phase diagram.

\begin{figure}[h]
\centering
\includegraphics[width=3.375in]{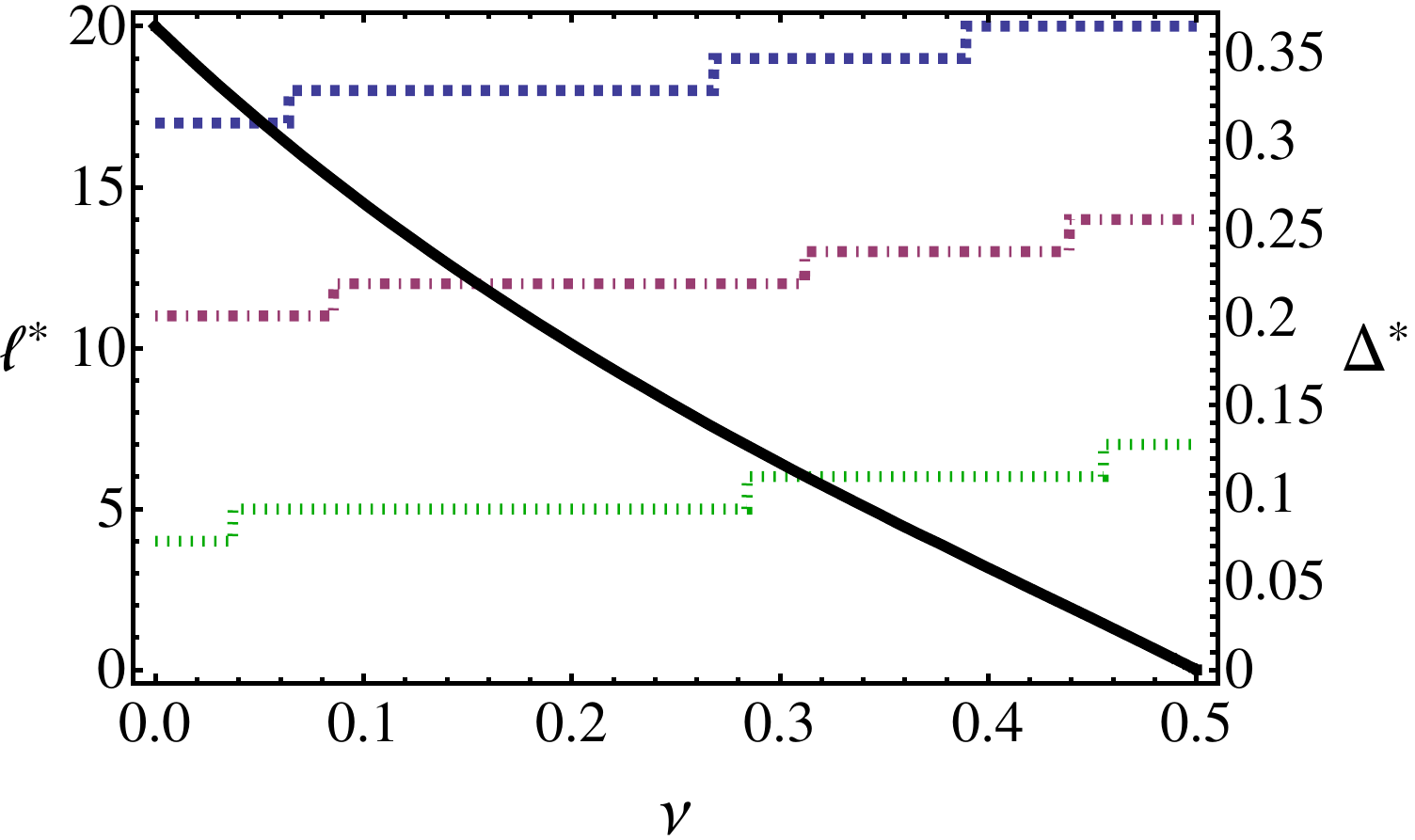}
\caption{
(color online) Plot of $\ell^{\star}$, the angular harmonic of the minimum
energy state of the sphere as a function of $\nu$
for $\alpha =100, 800, 2500$: dotted (green), dash-dotted (purple), and dashed (blue)
respectively. As the interior is made less compressible, the optimally buckled
shape has lower symmetry, i.e., higher $\ell^{\star}$ (left axis).
Additionally, the critical fractional excess area $\Delta^{\star}$ (black line,
right axis) decreases continuously to zero as $\nu \rightarrow 1/2$, the
incompressible limit,  for fixed $\alpha = 575$.
}
\label{fig:lstar-vs-nu}
\end{figure}

The dependence of these results on the Poisson ratio is generically weak -- see
Fig.~\ref{fig:lstar-vs-nu} --  except in the singular limit of $\nu \rightarrow
1/2$ where the material becomes incompressible. As $\nu$ approaches $1/2$, the symmetry of the optimally buckled state is gradually
reduced. A more incompressible interior results in more wrinkles for the buckled
state.  As expected, the energy cost of the isotropically expanded sphere
increases without bound as the sphere's interior is made more incompressible.
Consequently, the critical value of the excess fractional area $ \Delta $ decreases
continuously with increasing $\nu$. Finally at $\nu =1/2$ the isotropic solution is
disallowed by the compressibility condition and buckling occurs for all $\Delta
>0$ and all $\alpha$.

%%%%%%%%%%%  FIGURE
%%%%%%%%%%%%%%%

\begin{figure}[h]
\centering
\includegraphics[width=3.375in]{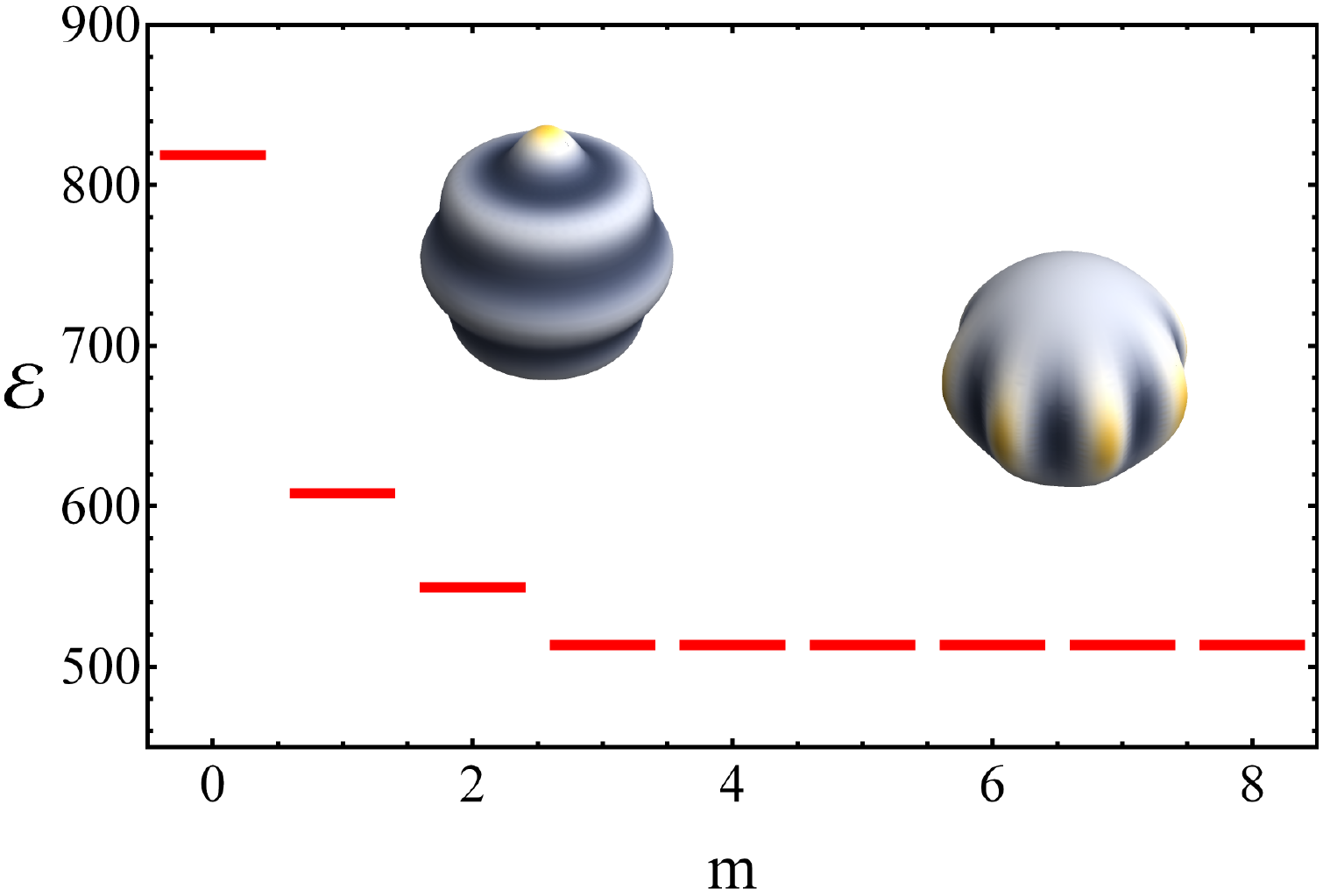}
\caption{(color online) Example of the degeneracy of the buckled state energy with respect to
$m$ for a representative value of $\ell = 8$ and $\alpha=575, \nu=0.2$ and
excess fractional area of $\Delta = 0.5$.  For fixed $\ell$ the energy initially decreases with $m$ until $m =3$.
All states with $m \ge 3$ are generate. Superposed on the plot are pictorial representations of the higher
energy state with $m=0$ (left) and one of the lowest energy states with $m = \ell =8$ (right).}
\label{fig:E-vs-m}
\end{figure}

When examining the energy of various buckled states for the same $\ell$, one
notes that it is a non-increasing function of $m$.  The buckled states having
rings of equal radial deformation extending azimuthally around the sphere, e.g.,
the $m=0$ state is always the highest energy state with the subspace
of equal $\ell$ buckled configurations. As $m$ is increased, the energy initially
decreases within increasing $m$, but reaches a plateau for $m \ge 3$ independent
of $\ell$ for $\ell >3$.  An example of this degeneracy is shown in
Fig.~\ref{fig:E-vs-m} for the $\ell =8$ with the same combination of $\alpha,
\nu$ as used in the previous figures. The fractional excess surface area is
fixed at $\Delta =0.5$.

%%%%%%%%%%%  FIGURE
%%%%%%%%%%%%%%%

%%%%%%%%%%  FIGURE
%%%%%%%%%%%%

This degeneracy can be understood as follows. Referring to Appendix B, we see
that the displacement field in the sphere's interior associated with a surface
deformation  proportional to $Y_{\ell m}$ contains terms with angular dependence
given by a sum of spherical harmonics with $m-1,m,m+1$. The elastic energy
density requires one more spatial gradient of  ${\bf u}({\bf x})$. Since the
gradient is a vector operator, the resulting elastic energy density will contain
a sum of terms having magnetic quantum numbers of $m-2, \ldots, m+2$. Recall
also our spherical harmonics $Y_{\ell m}$ are superpositions of complex
spherical harmonics with $\pm m$ so that the volume integral of the bulk energy density contains
an array of products of terms having magnetic quantum numbers between $-m -2,
\dots -m+2$ and $m-2, \ldots m+2$.  Cross products between these two sets of
terms vanish upon integration unless the sum of their magnetic quantum numbers
vanishes. There are more such nonvanishing products, however, if these two sets
of terms have an overlap; in other words if $m \le 2$, then there is an
accessible solution to $- m + 2 = m - 2$ and more products are nonvanishing in
the computation of the internal energy of the sphere. These terms necessarily
make a positive contribution to the elastic energy of the sphere since every
deformation adds to the elastic energy. Thus as $m$ is increased, the number of
these cross terms is reduced as and the total elastic energy of deformation
decreases. Once $m \ge 3$, however, there are no more cross terms.
The higher $m$ deformations are now degenerate, at least within the linear
elastic theory. Nonlinear elastic terms in the sphere's interior or surface
compression energies (we include only bending) probably lift this degeneracy,
but such effects are beyond the scope of the current analysis.

%%%%%%%%%%%  FIGURE
\begin{figure}[h]
\centering
\includegraphics[width=3.375in]{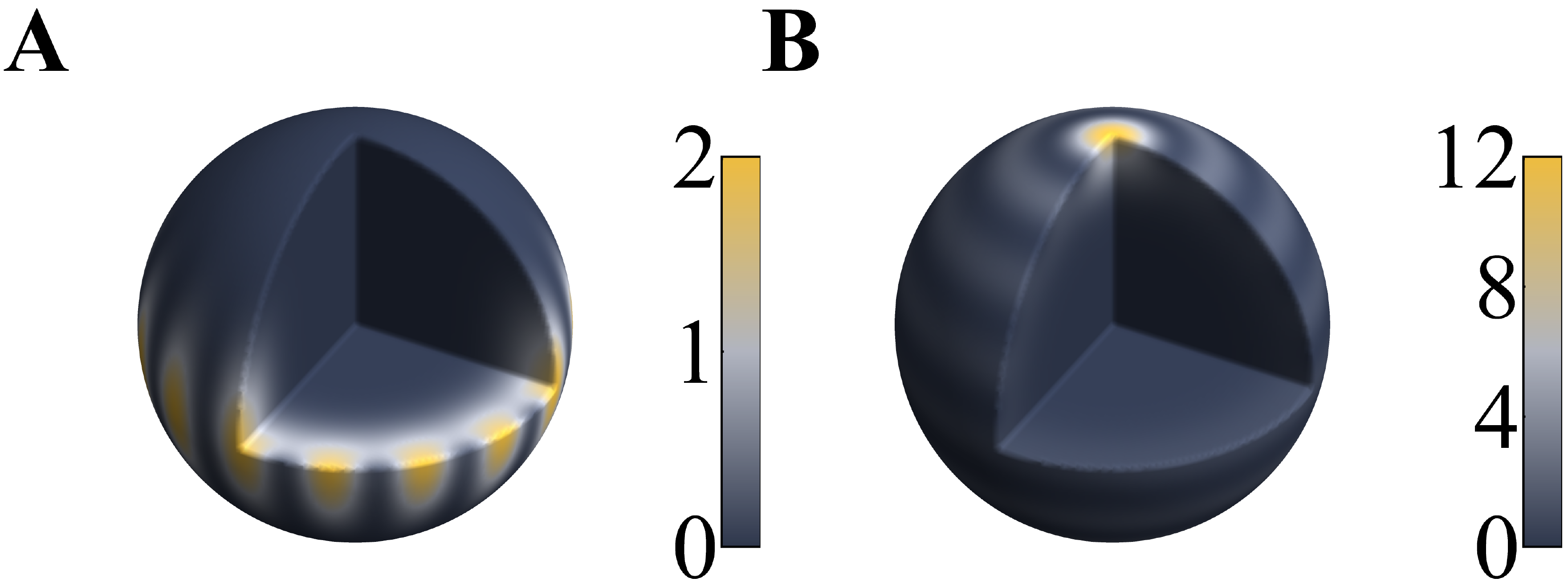}
\caption{
(color online) A comparison of the distribution of the elastic energy density in
spheres having the same fractional excess area $\Delta = 0.5$, and same elastic
parameters $\alpha = 575, \nu = .2$. On the left (A)  the excess
area is put into the $\ell =8, m=8$ mode. Here the energy density is primarily
spread out over the equator. On the right (B) the excess area is put into
the $\ell =8, m=0$ mode so that the sphere's surface has undulatory rings that
run along lines of constant latitude. The overall elastic energy of A is less than B --
note the difference in the energy scale associated with the two color maps.
}
\label{fig:mode-comparison}
\end{figure}
%%%%%%%%%%%%

The physical effect of these extra terms in the energy density is well
illustrated by plotting the distribution of elastic energy storage in two
spheres with $\ell =8$ surface deformations. Their energy densities are shown as
a heat map in Fig.~\ref{fig:mode-comparison} for the $m=0$  (left) and the $m=8$
(right) solutions. The $m=0$ mode is clearly higher in energy as can be seen by
noting the difference between the two color scales. From a comparison of the two
figures, it is clear that in both cases the elastic deformation energy is
confined to a boundary layer near the sphere's surface, as expected based on the
heuristic calculation. In the $m=0$ solution, however, there is a distinct
concentration of the elastic energy near the sphere's poles. This is due to the
fact that $m=0$ distortion lay along lines of longitude. Near the poles these
undulatory rings of deformation become tightly wrapped.  In the $m=8$
solution the elastic energy is spread out primarily over the equator of the
sphere, resulting in a lower total energy.

\begin{figure}[h]
\centering
\includegraphics[width=3.375in]{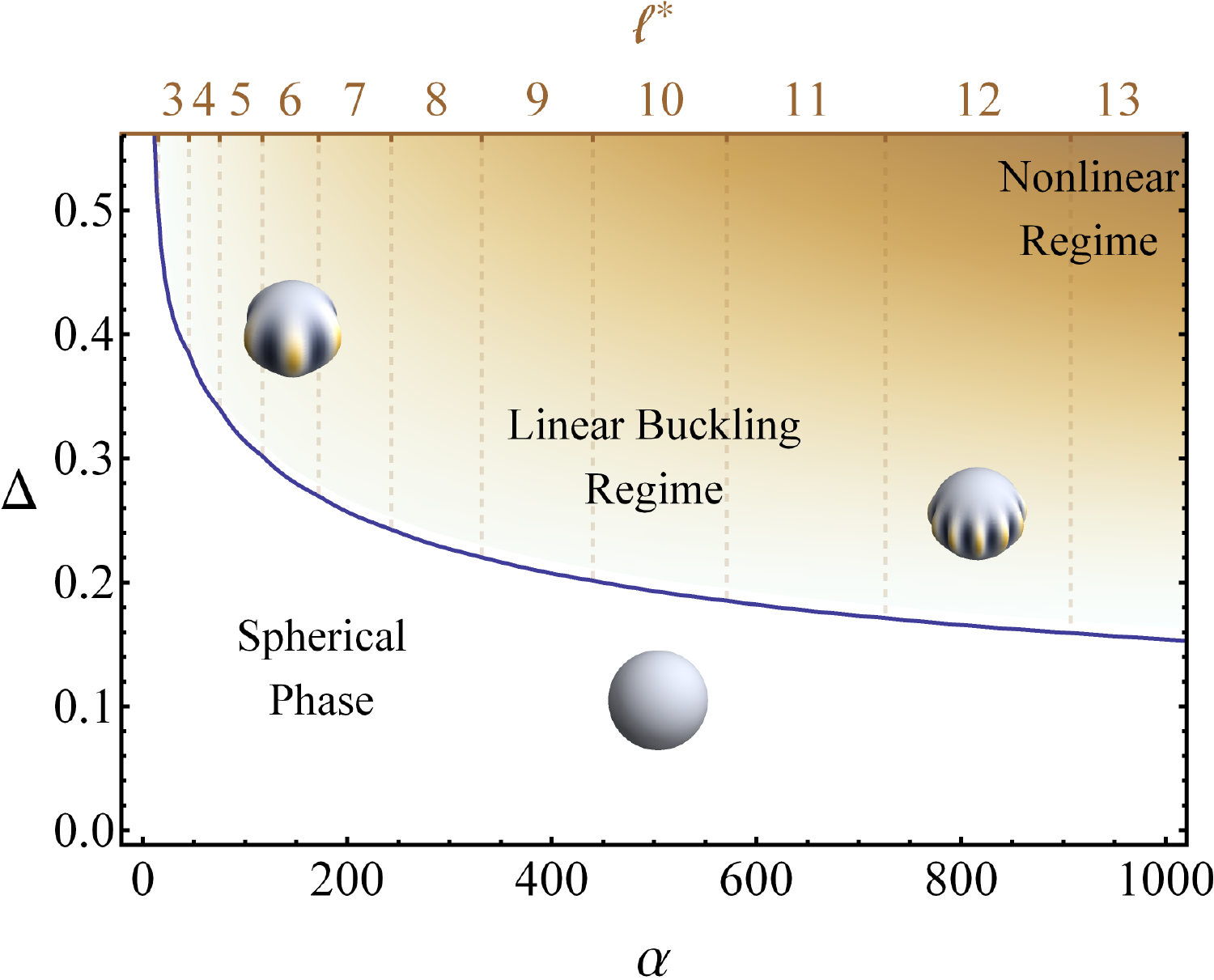}
\caption{(color online) Buckling phase diagram of core-shell structures spanned by the fractional excess area $\Delta$
and the dimensionless control parameter $\alpha$. The (solid, black) curve separates isotropically expanded
spheres from lower symmery buckled states. As $\alpha = \mu R^{3}/\kappa$
is increased the interior material stiffens relative to the bending modulus of the boundary, leading to
buckling at smaller $\Delta$.  Also, as $\alpha$ is increased the $\ell^{\star}$ of the buckled
state near the transition also increases -- see the top axis. Representative shapes for
these minimum energy buckling spheres are shown for $\ell^{\star} = 6,12$.}
\label{fig:phase-diagram}
\end{figure}

Putting these results together into a more unified picture, we present a phase
diagram for spherical elastic core/membrane shell structures in Fig.~\ref{fig:phase-diagram}.
The  diagram is spanned by the excess fractional area $\Delta$
and the control parameter $\alpha$ for a fixed value of the Poisson ratio $\nu =
0.2$.  This phase diagram is representative of other $\nu$ values with the exception of
the $\nu=1/2$ incompressible case, as mentioned above. The solid black in
Fig.~\ref{fig:phase-diagram} line divides the the phase space of such elastic
objects into
%%%%%%%%%%%  FIGURE
%%%%%%%%%%%%
isotropic and expanded spheres below line and lower symmetry buckled ones above
it. One  also observes the dependence of the critical fractional
excess area on $\alpha$.  As $\alpha$ is increased $\Delta^{\star}$ indeed
decreases and the symmetry of the buckled state decreases as well. The angular
harmonic of the optimal buckled state $\ell^{\star}$ is shown by the horizontal
axis along the top of the figure.  The representative figures show the $\ell =m$
state, which, as described earlier, is always part of the degenerate subspace of
optimally buckled shapes for $\ell^{\star} \ge 3$. Such buckled shapes are qualitatively
similar to the buckled gel-filled vesicles of \cite{Finan2009a, Viallat2004}.

\begin{figure}[h]
\centering
\includegraphics[width=3.375in]{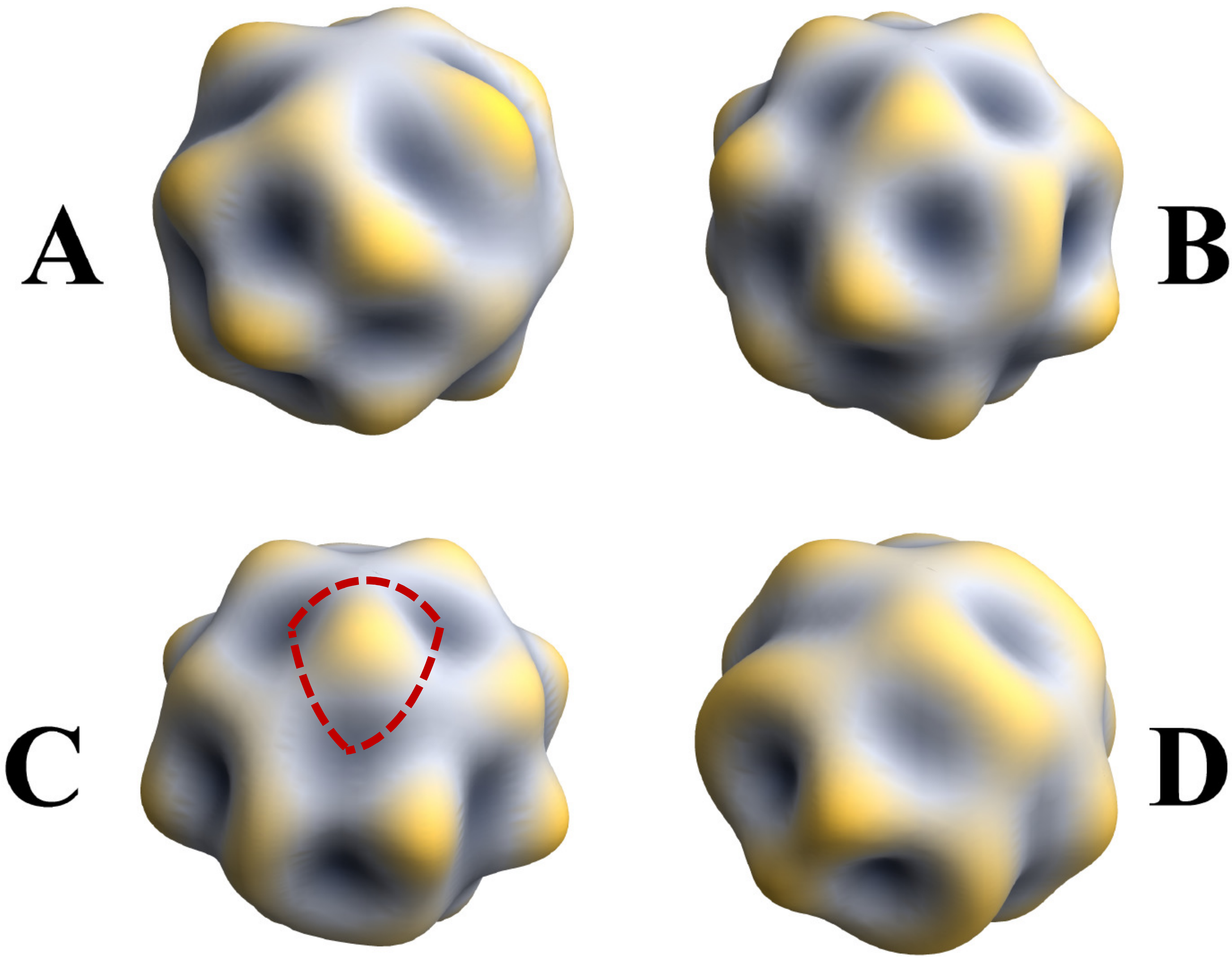}
\caption{(color online) Examples of mixing the degenerate buckling modes for a system with
 $\mu = 0.2$ and  $\alpha=300$. For this value of the control parameter, the degenerate modes are
the $\ell=8$ and $m\in[3,8]$. Each example is an equally weighted mixture of buckling modes with amplitude such that
$\Delta = 0.4$.  \textbf{A}: $m=4\mbox{ and }6$;
\textbf{B}: $m=3 \mbox{ and }7$; \textbf{C}:  $m=3\mbox{ and }8$; \textbf{D}:
$m=3,\ 5,\ 7,\mbox{ and }8$. }
\label{fig:triangle}
\end{figure}

There are a variety of other buckled core shell structures.  In particular, Cao and coworkers
investigated the buckling of Ag core/ SiO$_{2}$ shell particles experimentally~\cite{Cao2008}
and numerically~\cite{Li2011}, where, relying the materials'
differential thermal expansion, buckling was induced by temperature
change. In these systems, a triangular pattern of ridges where observed near the transition from
the smooth spherical state, while, deeper the buckled phase, labyrinthine patterns were observed. The
triangular patterns observed near the transition have lower symmetry than an single spherical harmonic mode
predicted by the linear theory. Such triangular patterns, however, can be simply reproduced from out results by
forming a linear superposition of the degenerate buckling modes predicted herein. This is shown in Fig.~\ref{fig:triangle}
where we reproduce examples of lower symmetry buckling states by combining degenerate buckling modes all with $\ell =8$.
One generically finds a triangular pattern of dimples in such cases, as shown by the dashed (red) line in
Fig.~\ref{fig:triangle}~C. Understanding whether these mixed modes truly represent lower energy solutions
or are the result of some underlying quenched disorder in the core/shell system is beyond the scope of our linear
elasticity approach. Nevertheless, from our analysis it is easy to see that either small elastic nonlinearities or disorder may
select such mixed modes from the space of degenerate states that we identify, and that these mixed mode solutions are
qualitatively similar to those observed.
%%%%%%%%%%%  FIGURE
%%%%%%%%%%%%

\section{Summary}
\label{sec:summary}

We have demonstrated the analog of Euler buckling in spherical core-shell
structures where the shell is treated as a tensionless lipid membrane, i.e., one
whose elasticity is controlled solely by a bending modulus, $\kappa$.  Such
systems are found in gel-filled vesicles, but with some modifications to the
treatment of the surface elasticity of the bounding membrane (e.g., adding an
area compression modulus) these calculations have more general relevance to a
variety of soft elastic solids coupled to thin shells, such any number of
colloidal core/shell particles~\cite{Rowat2005,Rowat2006,Yin2008,Li2011,Shum2011}, or even the blastocyst  that
transitions from spherical to ovoid shape during embryonic development~\cite{Mattson1990}.
Following our exploration of a heuristic scalarized
elasticity theory, we identified one control parameter for the transition upon
increasing the bounding surface area between an isotropically expanded sphere
and a buckled one.  This control parameter $\alpha = \mu R^{3}/\kappa$ measures
the relative compliance of the sphere's interior to its boundary. In the full
elastic theory presented above, there is also a continuous and weak dependence
of the results on the Poisson ratio of the elastic interior, but as long as the
material remains somewhat compressible, these effects are subdominant.

It is of interest then to determine the expected magnitude of $\alpha$ for
various systems of current interest. To do this it convenient to rewrite this
dimensionless quantity as the ratio of two lengths: $\alpha = R^{3}/\xi^{3}$,
where the numerator is of purely geometric origin -- the size of the sphere --
and the denominator $\xi = (\kappa/\mu)^{1/3}$ is a length set by the ratio of
the surface to bulk moduli of the system.  For many soft polymeric gels and even
for the interior of eukaryotic cells the typical values of the shear modulus are
in the range of $10^{2}$--$10^{4}$Pa. Bending moduli of lipid bilayers are
commonly found to be in the range of 10 -- 100 $k_{\rm B}T$ at room temperature.
This range of values generates a natural length scale of $\xi \sim
10^{-2}$--$10^{-1}\mu$m.  Thus, for spheres of typical colloidal dimensions,
$R\sim 1-10 \mu m$, we find a range of $\alpha \sim 10^{3} - 10^{9}$, justifying
our focus on the large $\alpha$ regime. Since the optimally buckled state
spherical harmonic $\ell^{\star}$ scales as $\alpha^{1/3}$, we predict that all
buckled soft spheres of micron size will have $\ell^{\star} \sim {\cal O}(10)$.
Such wrinkled states are indeed observed in the gel-filled
vesicles~\cite{Viallat2004}.  We imagined that one
encounters the mechanical instability leading to buckling via increasing
the surface area of the bounding membrane at fixed interior volume. The gel-filled
vesicle experiments generate the instability by reducing the interior at fixed
membrane surface area. Within linear elasticity theory as used here, these two approaches to the
instability are equivalent. There may be biophysical examples of the same mechanical instability
driven by membrane growth occurring in cells, nuclei, and, more broadly, during development.

 It should be possible to extend this analysis to consider the dynamics morphological change
 in growing systems. We identify in our present work the requisite excess area for the
 morphological transition. As long as stress relaxation is fast compared to growth, our analysis remains
 valid for any growth model: morphology is simply slaved to excess area.  However, if one were
 to consider the interior to be viscoelastic with a sufficiently long stress relaxation time, then one should
 encounter interesting new dynamics of the instability by
explicitly including material growth as explored in Refs.~\cite{BenAmar2005,Goriely2005}.

A similar buckling transition has been analyzed for the geometry
of a thin shell or membrane on a flat elastic substrate~\cite{Chen2004}. In
that work, the authors describe a critical buckling stress as opposed to
a critical excess surface area, as we do. The two are simply related in linear
elasticity theory. We find that our results may be equivalently described in
terms of a critical stress in the membrane for buckling. The key distinction
between buckling on a sphere and on a flat substrate is that the sphere's
radius now enters the critical stress. We may, in fact, express that critical
stress in terms of the control parameter $\alpha$ and a dimensionless
function of the Poisson ratio of the internal material. The result is given in
Appendix C.

Extending the above analysis to core-shell structures where the shell also
contains a shear and area compression modulus remains an open and quite complex
problem. We do expect the general phenomenology observed here to remain valid.
Specifically, surface buckling will occur at a critical excess area and high
moduli elastic interiors will push this morphological transition to lower
critical fraction excess areas and require progressively lower symmetry buckled
states for areas just above the critical value. We expect that the additional
moduli associated with the surface will effectively stiffen the shell towards
buckling and serve to further stabilize the isotropically extended sphere to
larger values of $\Delta$ than those found here.

The theory presented here applies only for small distortions where linear
elasticity remains valid.  For the expected levels of fractional excess area
$\Delta^{\star} \sim 0.5$ the typical size of the radial displacements at the
surface are on the order of $u_{r}/R \sim 0.05$, so that it seems reasonable to
suppose that the application of the theory is valid up to the predicted
morphological transition.  Clearly as more area is added beyond that point,
nonlinear effects will become relevant. Numerical simulations~\cite{Li2011} of
elastic core-shell structures under ``drying,'' where volume is removed from the
interior rather than area added to the surface, have been performed. The results
suggest that patterns reminiscent of those predicted here near the transition
evolve under the further reduction of interior volume towards deeply folded
surfaces where the folds meet in three-fold junctions that roughly tessellate the
surface. This suggests that elastic nonlinearities neglected here lead to fold
formation, as might be expected based on the extensive work on folding of
two-dimensional globally elastic~\cite{Cerda2003} and flat elastica coupled to
elastic or viscous subspaces~\cite{Diamant2011}. Exploring the folding
phenomenon and the interactions of folds on a curved surface remains an open
problem.

\begin{acknowledgments}
AJL acknowledges enjoyable conversations with A.A. Evans and W.S. Klug. He also
acknowledges partial support from NSF Grant no. DMR-1006162.
\end{acknowledgments}

\appendix

\section{Modeling growth via applied stress}
In our calculation we impose the excess area of the sphere's surface by applying
constraint surface tractions as required to generate a specified radial
displacement of the surface. The amplitude of that radial displacement is then
determined by desired excess area. It is reasonable to ask whether such a
procedure is equivalent to demanding the growth of the sphere's surface and then
computing the elastic distortions required to accommodate that growth. In this
appendix we demonstrate that, within a linear elastic theory, there is a simple
correspondence between changes in the elastic reference state, reflecting
material growth, and the application of constraint tractions at the boundary of
the elastic object.

To begin we consider the simplest case of one elastic degree of freedom.
Imagine a single particle bound to the origin by a spring having spring constant
$k$. Its energy of deformation in one dimension is
\begin{equation}
\label{spring}
E = \frac{1}{2}k x^{2},
\end{equation}
where $x$ is the displacement of the particle. In this simple example, a change
of reference state entails shifting the origin of the spring from $x=0$ to
$x=x_{0}$, yielding an energy function
\begin{equation}
\label{shifted-spring}
E = \frac{1}{2}k (x-x_{0})^{2} = \frac{1}{2}k x^{2} - F_{0}x + E_{0},
\end{equation}
where $F_{0} = k x_{0}$ and the constant in the energy $E_{0} = \frac{1}{2}k
x_{0}^{2}$ has no effect on the statics or dynamics of the system.  From
Eq.~\ref{shifted-spring} we see that the effect of shifting the reference state
is equivalent to adding a constraint force $F_{0}$ to the original energy
function for the unshifted spring in Eq.~\ref{spring}.

We now consider a simplified one-dimension version of linear elasticity theory.
We define the reference space labeling the undeformed mass points of the system
by $X$ on a segment of the $x$-axis in the range $[0,1]$. We now grow this
elastic material via the transformation $X \rightarrow \lambda X$, with $\lambda
= 1 + \delta > 1$ for growth. In the original system displacement vectors $u =
{\bf u}\cdot \hat{x}$ are defined in terms of the location of those same mass
points in the deformed state by
\begin{equation}
u(X) = R(X) - X,
\end{equation}
where the mapping $X \rightarrow R(X)$ defines the deformation state of the
one-dimensional elastic continuum. As expected from the global translational
invariance of the energy, the energy density $e(X)$ of the elastic body is
proportional to derivatives of the displacement vector with respect to the {\em
reference state label} $X$. In the system with growth, this becomes

\begin{equation}
e(X)  = \frac{1}{2} \mu \left( \frac{\partial }{\partial X} \left[ R(X) - X  - \delta X \right]  \right)^{2},
\end{equation}
where we have introduced a single elastic constant $\mu$.

Expanding the square as we did for the simple spring system we find that the
elastic energy density of deformation in the system with growth $\lambda$ is
given by
\begin{equation}
e(X)  = \frac{1}{2} \mu \left( \frac{\partial u}{\partial X} \right)^{2} - \sigma \frac{\partial u}{\partial X} + e_{0},
\end{equation}
where $\sigma = \mu \delta $ now acts as a surface traction and the constant in
the energy density is, once again, physically irrelevant. Thus, we see in a one dimensional model of linear elasticity, growth of
the reference state is identical mechanically to applying the appropriate set of
surface tractions. In our calculation, we did not know these tractions {\em a
priori} , but determined them later by required a fixed increase in the surface
area of the growing and elastically deformed structure.

The principle demonstrated here applies for three dimensional vectorial
displacements at the cost of some trivial complexity involving indices. Note
that the equivalence exploited here is valid only if one assumes a linear elastic
material for which all terms in the energy functional are quadratic in the
strain tensor. It is only in this case that expanding the square generates a
linear shift in the energy interpreted as the surface traction and an irrelevant
additional constant. Future studies of nonlinear elastic systems will require a
more sophisticated approach to the problem of changing the reference state of
the material.

\section{Solving for the elastic deformation of the sphere's interior}

In this appendix we examine in more detail the solution to elasticity problem proposed in the text. Namely, if one were to specify a radial displacement of
the surface of an elastic sphere in terms of a single spherical harmonic, what is the form of the displacement field in the sphere's interior? We solved this
problem by asserting that one may find a harmonic vector field ${\bf S}_{\ell m}$ that satisfies the required radial displacements on the sphere's surface.  In this
appendix we present some of the details of that computation and give the solution of the displacement field ${\bf u}({\bf x})$.

On the surface of the sphere we require the vector field ${\bf S}_{\ell m}$ to be given by
\begin{equation}
{\bf S}_{\ell m} = g_{\ell m} \hat{r} Y_{\ell m}.
\end{equation}
To construct a harmonic vector is of this form we write the Cartesian components of this vector field at the surface, expand these components
in sums of spherical harmonics, and then multiple each of them the appropriate factor of $(r/R)^{\ell}$ to make these function harmonic in the
interior. These $x,y,z$ Cartesian components are then of the form  $Y_{\ell m} \sin (\theta) cos(\phi)$,
$Y_{\ell m} \sin (\theta) cos(\phi)$, $Y_{\ell m} \cos \theta$ respectively. Using the usual rules for the addition of angular momentum, these products
can be expanded in sums of other spherical harmonics. Specifically, we may use the fact that the product of two spherical harmonics may be expanded
in terms of a sum of spherical harmonics using the Wigner $3-j$ symbols
\begin{eqnarray}
Y_{\ell_1}^{m_1}Y_{\ell_2}^{m_2}&=& \sum_{\ell m}\sqrt{\frac{(2 \ell_{1}+ 1)(2 \ell_{2}+1) (2 \ell +1)}{4 \pi}} \times \notag \\
 & & \hspace{-0.7cm} \times \left( \begin{array}{ccc}
 \ell_{1} & \ell_{2} & \ell\\
 m_{1} & m_{2} & m
 \end{array} \right) Y_{\ell}^{-m}
  \left( \begin{array}{ccc}
 \ell_{1} & \ell_{2} & \ell \\
0 & 0 & 0
 \end{array} \right).
\end{eqnarray}
The Wigner $3-j$ symbols are simply related to the better known Clebsch-Gordan coefficients. The reader may refer to any standard reference
on quantum mechanics~\cite{Condon-Shortley:51,Messiah1999}.  In the products above $\ell_{1}, m_{1}$ refer to the angular harmonic of the
imposed surface deformation and $\ell_{2}, m_{2}$, the decomposition of the radial unit vector into its Cartesian components.Thus,
$\ell_{2} =1$, and $m_{2} = \pm 1, 0$. Given this simplification, it is useful to define $A_{\pm}^{\pm,0}$ where the lower indices
refer an increase or decrease of the total angular momentum: $\ell \rightarrow \ell \pm 1$, while the upper index refers to the change in its $z$-axis
projection: $ m \rightarrow m \pm1, m $
\begin{equation}
\label{A-def--}
A_{-}^{\pm}(\ell,m) = \frac{-1}{2}\sqrt{\frac{3}{2\pi}}
\sqrt{\frac{(\ell\mp m-1)(\ell\mp m)}{(2\ell-1)(2\ell+1)}}
\end{equation}
\begin{equation}
\label{A-def-+}
A_{+}^{\pm}(\ell,m) = \frac{1}{2}\sqrt{\frac{3}{2\pi}}
\sqrt{\frac{(\ell\pm m+2)(\ell\pm m+1)}{(2\ell+1)(2\ell+3)}}
\end{equation}
for the terms generating $m$ values of $m\pm1$ and
\begin{equation}
\label{A-def-0-}
A_{-}^{0}(\ell,m) = \sqrt{\frac{3}{4\pi}}
\sqrt{\frac{(\ell^2-m^2)}{(2\ell-1)(2\ell+1)}}
\end{equation}
\begin{equation}
\label{A-def-0+}
A_{+}^{0}(\ell,m) = \sqrt{\frac{3}{4\pi}}
\sqrt{\frac{(\ell+ m+1)(\ell- m+1)}{(2\ell+1)(2\ell+3)}}
\end{equation}
for the terms with the same $m$ value as the radial surface deformation.

At this point it is useful to define $Y_{\ell m}$ for $m<0$
\begin{equation}
Y_{\ell m}= \tfrac{1}{i\sqrt{2}}\left[Y_\ell^{|m|}-(-1)^mY_\ell^{-|m|}\right]\quad m<0
\end{equation}
This is simply a rotation of $Y_{\ell |m|}$ mode through an angle $\phi_0 = \tfrac{\pi}{2m}$.
Defining $Y_{\ell m}$ for all $m \in  \mathbb{I}$ allow us to describe a mode rotated
by any angle. This notation is useful in the Cartesian representation of the deformation field.

Using this decomposition of the ${\bf S}_{\ell m}$ vector into its spherical harmonic components, multiplying each by the appropriate power $(r/R)^{\ell}$
to make this vector field harmonic in the sphere's interior, and using the above notation, we find:
\begin{eqnarray}
\nonumber
S_x&=&\mbox{Sgn}(m)g_{\ell m}\sqrt{\frac{2\pi}{3}}  \left\{  \left(  \frac{r}{R}\right)^{\ell -1} \left[ A_{-}^{-}(\ell,m) \times \right. \right.\\
\nonumber
& & \hspace{-1.2cm} \left. \left(1-2\delta_{m,0}\right) Y_{\ell-1,m-1}  - A_{-}^{+} (\ell,m) \left(1-\delta_{m,-1}\right) Y_{\ell-1,m+1} \right] \\
\nonumber
& & \hspace{-1.2cm} + \left(\frac{r}{R}\right)^{\ell +1} \left[ A_{+}^{-}(\ell,m)  \left(1-2\delta_{m,0}\right) Y_{\ell+1,m-1}  \right. \\
\nonumber
& & \hspace{-1.2cm}  - \left. \left. A_{+}^{+} (\ell,m) \left(1-\delta_{m,-1}\right) Y_{\ell+1,m+1}  \right] \vphantom{\left(\frac{r}{R}\right)^{\ell -1}} \right\}\\
\end{eqnarray}
\begin{eqnarray}
\nonumber
S_y&=&-\mbox{Sgn}(m)g_{\ell m}\sqrt{\frac{2\pi}{3}}  \left\{  \left(  \frac{r}{R}\right)^{\ell -1} \left[ A_{-}^{-}(\ell,m) \times \right. \right.\\
\nonumber
& & \hspace{-1.2cm} \left. Y_{\ell-1,-m-1}  + A_{-}^{+} (\ell,m) Y_{\ell-1,-m+1} \right] + \\
\nonumber
& & \hspace{-1.2cm} + \left(\frac{r}{R}\right)^{\ell +1} \left[ A_{+}^{-}(\ell,m)  Y_{\ell+1,-m-1}  \right. \\
\nonumber
& & \hspace{-1.2cm}  - \left. \left. A_{+}^{+} (\ell,m)  Y_{\ell+1,-m+1}  \right] \vphantom{\left(\frac{r}{R}\right)^{\ell -1}} \right\}\\
\end{eqnarray}
\begin{eqnarray}
\nonumber
S_z& = & g_{\ell m} \sqrt{\frac{4\pi}{3}}  \left\{  \left(  \frac{r}{R}\right)^{\ell -1}  A_{-}^{0}(\ell,m) Y_{\ell-1,m}   + \right. \\
& & \left. \left(\frac{r}{R}\right)^{\ell +1}  A_{+}^{0}(\ell,m) Y_{\ell+1,m} \right\},
\end{eqnarray}
where we have written the result in terms of the real spherical harmonics defined in the text so that each term of the above harmonic
vector is explicitly real.  Finally, we have introduced {\em sign function} $\mbox{Sgn}(x) = +1$ for nonnegative $x$ and $-1$ otherwise.

Finally, for completeness we record the Cartesian components of the displacement vector field ${\bf u}({\bf x})$ for a given radial deformation field on
the sphere's surface taken to be a single (real) spherical harmonic:

\begin{widetext}
\begin{equation}
\begin{aligned}
u_x&=g_{\ell m}\frac{r^{\ell-1}}{4R^{\ell+1}}
\left[\left(2 r^2 \sqrt{\frac{1}{\left(4 \ell^2-1\right)}}-
\frac{\left(R^2-r^2\right) (\ell-4 \nu +5)
\sqrt{\frac{(2 \ell+1)}{(2 \ell-1) }}}{\ell (4 \nu -3)+2 \nu -1}
\right)\right.\\
&\left(\sqrt{(\ell-m-1) (\ell-m)}Y_{\ell-1,m+1}-\mbox{Sgn}(m-1)^{m-1}
\sqrt{(\ell+m-1)(\ell+m)}Y_{\ell-1,\left| m-1\right|}\right)\\
&\left.\vphantom{\frac{\sqrt{\frac{(2 \ell+1) (\ell+m-2)!}{(2 \ell-1)(\ell-m)!}}}
{\ell (4 \nu -3)+2 \nu -1}}
-2 r^2   \sqrt{\frac{2\pi}{3}}
\left(A_{+}^{+} (\ell,m) Y_{\ell+1,m+1} -\mbox{Sgn}(m-1)^{m-1}\  A_{+}^{-} (\ell,m) Y_{\ell+1,\left| m-1\right|}\right)
\right]
\end{aligned}
\end{equation}

\begin{equation}
\begin{aligned}
u_y&=g_{\ell m}\frac{r^{\ell-1}}{4R^{\ell+1}}
\left[\left(2 r^2 \sqrt{\frac{1}{\left(4 \ell^2-1\right)}}-
\frac{\left(R^2-r^2\right) (\ell-4 \nu +5)
\sqrt{\frac{(2 \ell+1)}{(2 \ell-1) }}}{\ell (4 \nu -3)+2 \nu -1}
\right)\right.\\
&\left(\sqrt{(\ell-m-1) (\ell-m)}Y_{\ell-1,m-1}+\mbox{Sgn}(m-1)^m
\sqrt{(\ell+m-1) (\ell+m)}Y_{\ell-1,-\left| m+1\right|}\right)\\
&\left.\vphantom{\frac{\sqrt{\frac{(2 \ell+1) (\ell+m-2)!}{(2 \ell-1)(\ell-m)!}}}
{\ell (4 \nu -3)+2 \nu -1}}
-2 r^2   \sqrt{\frac{2\pi}{3}}
\left(A_{+}^{-} (\ell,m)Y_{\ell+1,m-1} +
\mbox{Sgn}(m-1)^m\  A_{+}^{+} (\ell,m)Y_{\ell+1,-\left| m+1\right|}\right)
\right]
\end{aligned}
\end{equation}

\begin{equation}
\begin{aligned}
u_z&= g_{\ell m}\frac{r^{\ell-1}}{R^{\ell+1}}\sqrt{\frac{4\pi}{3}}
\left[r^2   A_{+}^{0} Y_{\ell+1,m} +A_{-}^{0}
\frac{\left(2 \ell^2+5 \ell+3\right) r^2-(2 \ell+1) R^2 (\ell-4 \nu +5)}{2 (\ell (4 \nu -3)+2 \nu -1)} Y_{\ell-1,m}
\right].
\end{aligned}
\end{equation}
\end{widetext}
This completes the analysis of the displacement field inside the elastic sphere
when a radial deformation whose amplitude is proportional to a single spherical
harmonic is imposed on the surface.

\section{Critical Stress for Buckling}

One may calculate the critical surface stress required for the
morphological transition using the elastic theory. We calculate the critical
stress by computing the surface stress of the $\ell=0$ mode as a function of
$g_{00}$. We then use the critical fractional excess area $\Delta^\star$ to
solve for the value of $g_{00}$ at the buckling transition.

The critical stress may be
written in terms of the control parameter $\alpha$, $\kappa/R^{3}$, and a dimensionless
function of $\nu$, the Poisson ratio of the interior. This function $f(\nu)$ is difficult to produce in 
closed form, but we find that the following polynomial serves as a good numerical approximation 
for $\nu \ge  0$: 
\begin{equation}
f(\nu) = -7.92208 \nu ^3+10.2125 \nu ^2-9.19217 \nu +3.0332.
\end{equation}
In terms of this function the critical stress $\sigma^\star$ is given by
\begin{equation}
	\sigma^\star= \frac{\kappa}{R^3}\frac{(1+\nu) \sqrt{2/\pi}}{1-2\nu}
	\left(\frac{f(\nu)\alpha^{2/3}}{2}-\frac{f^2(\nu)\alpha^{1/3}}{8}\right).
\end{equation}

\bibliography{WBC_paper-Citations}
\end{document}